\documentclass[aps,preprint,preprintnumbers,nofootinbib,showpacs]{revtex4}

\usepackage{amsmath}
\usepackage{amssymb}
\usepackage{mathtools}
\usepackage{dsfont}
\usepackage{color}
\usepackage{graphicx,accents}
\usepackage[normalem]{ulem}

\def\bea{\begin{eqnarray}}
\def\eea{\end{eqnarray}}
\def\sea{\nonumber \\&&}
\def\lla{\left\langle}
\def\rra{\right\rangle}

\def\ssc{\scriptscriptstyle}
\def\lsim{\mathrel{\raise.3ex\hbox{$<$\kern-.75em\lower1ex\hbox{$\sim$}}} }
\def\gsim{\mathrel{\raise.3ex\hbox{$>$\kern-.75em\lower1ex\hbox{$\sim$}}} }

\begin{document}
\preprint{{\vbox{\hbox{NCU-HEP-k096}
\hbox{Jan 2023}
\hbox{rev. Mar 2023}
\hbox{ed. Apr 2023}
}}}
\vspace*{.3in}


\title{\boldmath  $E=mc^2$ versus Symmetry for Lorentz Covariant Physics
\vspace*{.1in}}

\author{Otto C. W. Kong$^{a,b}$ and Hock King Ting$^{a}$}

\affiliation{$^{a}$Department of Physics and Center for High Energy and High Field Physics,
National Central University, Chung-li, Taiwan 32054  \\
$^{b}$Quantum Universe Center, Korea Institute for Advanced Study,
  Seoul, 02455 Republic of Korea
\vspace*{.2in}}


\begin{abstract}
\vspace*{.1in}
The famous equation $E=mc^2$ is a version of particle mass being
essentially the magnitude of the (energy-)momentum four-vector in 
the setting of `relativistic' dynamics, which can be seen as dictated
by the  Poincar\'e symmetry adopted as the relativity symmetry.
However, as Einstein himself suggested, the naive notion of momentum 
as mass times velocity may not be right. The Hamiltonian formulation 
perspective gives exactly such a setting which in the case of motion 
of a charged particle under an electromagnetic field actually has 
the right, canonical, momentum four-vector with an evolving magnitude. 
The important simple result seems to have missed proper appreciation. 
In relation to that, we present clear arguments against taking the 
Poincar\'e symmetry as the fundamental symmetry behind `relativistic' 
quantum dynamics, and discuss the proper symmetry theoretical 
formulation and the necessary picture of the covariant Hamiltonian 
dynamics with an evolution parameter that is, in general, not a 
particle proper time. In fact, it is obvious that the action of any 
position operator of a quantum state violates the on-shell mass 
condition. The phenomenologically quite successful quantum field 
theories are `second quantized' versions of `relativistic' quantum 
mechanics. We present a way for some reconciliation of that with
our symmetry picture and discuss implications.
\\[.2in]

\end{abstract}

\maketitle

\section{Introduction}
Poincar\'e symmetry has been generally taken as the relativity
symmetry for Lorentz covariant physics, the so-called `relativistic' 
theories, quantum and classical. The related equation $E=mc^2$ 
may be the most popular physics equation, though the general
understanding of the simple equation has been loaded with many 
confusions \cite{O,H}, as would be expected. A key question we 
want to address here though, is if the `correct' picture about it, 
from the `standard'  `relativistic' theories, is indeed correct. 
The article launches a serious challenge to the question, at least 
for the case of quantum and classical particle dynamics, mostly 
from the symmetry theoretical formulation and the closely related 
Hamiltonian dynamics points of view. Implications for quantum
field theory would also be discussed.

The `standard'  `relativistic' theories are supposed to be 
obtainable, symmetry theoretically, from representations of 
the Poincar\'e symmetry. The irreducible representations of 
a symmetry, for the description of elementary systems 
(indecomposable into separated parts so long as the symmetry 
is concerned), are classified by its Casimir invariants. Note 
that we use the word representation in this article only in 
the straight symmetry theoretical sense, and talk about 
representations of a symmetry instead of a group or a Lie 
algebra, as a representation of a Lie algebra actually defines 
a representation of the Lie group, the universal enveloping 
algebra, and the group $C^*$-algebra as well as its various 
extensions all of which play some role in the physical theory. 
The Poincar\'e symmetry has as generators the six generators  
$J_{\mu\nu}$ of Lorentz symmetry as (spacetime) rotations 
and four generators $P_{\mu}$ as (spacetime) translations. 
$P_{\mu}P^{\mu}$ gives a Casimir invariant. For a particle, 
classical or quantum, the result says the magnitude square 
of the momentum four-vector, representation of $P^{\mu}$ as 
observables, has the value of $(-)m^2c^2$. That is called the 
on-shell mass condition. The reading of it in the particle 
rest frame gives $E=mc^2$. However, we are going to present 
arguments against seeing Poincar\'e symmetry as the right 
symmetry for the theories. The correct symmetry picture 
gives a different insight into the matter. 

It is interesting to note that though it was Einstein who first 
gave the $E=mc^2$ result and taught us that mass is a form of 
energy, he was {\em not} exactly in support of the unconditional 
on-shell mass condition, here as $p_\mu p^\mu = -m^2c^2$, for 
a (classical) particle \cite{H}. In a 1935  paper \cite{E}, 
Einstein actually raised the concern that his `momentum' 
four-vector as $m \frac{dx^\mu}{d\tau}$, $\tau$ being the particle 
proper time, may not be truly important conserved dynamical 
quantity in relation to the usual notion of conservation of 
momentum. From the symmetry theoretical formulation and the point 
of view of Hamiltonian dynamics, the latter is the canonical 
momentum as the representation of the $P^\mu$ generators of the 
symmetry. That is to say, Einstein thought that if the true 
energy-momentum four-vector $p^\mu$ is given by 
$m \frac{dx^\mu}{d\tau}$ and if the equation $p_\mu p^\mu = -m^2c^2$
is satisfied in a general situation was an open question.

Our focus is more on the quantum theories. It is well known 
that the usual theorists of `relativistic' quantum mechanics for 
particles of various spins, as presented in the textbooks, are of 
little practical applications. At least part of the reasons is 
related to the issues we are after here. That is a stunning 
contrast to the case of the `nonrelativistic' theory, which 
should be like its approximation. We have a nice way to look 
at Newtonian mechanics as an approximation to `nonrelativistic' 
quantum mechanics from a symmetry theoretical point of view, based 
on a scheme of symmetry contraction \cite{070}. Essentially, the 
contraction trivializes the Heisenberg commutation relation in the 
representation picture reducing the quantum to the classical. That
also has deformation quantization fitting in as the deformation of 
the observable algebra, say as the representation of the universal 
enveloping algebra, as a result of the deformation of the basic
symmetry Lie algebra. The deformation is the exact inverse of the 
contraction. The Poincar\'e symmetry does not allow any parallel. 
We have rather presented a new theory of Lorentz covariant quantum
mechanics \cite{087} that fit in well with all that, having
contractions that give the `nonrelativistic' or classical 
approximations correctly. That is based on the representation of
the $H_{\!\ssc R}(1,3)$ symmetry we discussed below. This article
helps to clarify why the latter symmetry, instead of the Poincar\'e 
symmetry, is the right one to be considered.

Most physicists are not much concerned about the unpleasant 
situation of `relativistic' quantum mechanics because the field
theory version seems to be working perfectly well, giving all 
the success in the description of results from experimental high 
energy physics. Yet, quantum field theories are supposed to be
only the `second quantized' version of `relativistic' quantum 
mechanics. If the latter has to be revised with the symmetry
picture, one has to worry if the same applies to quantum field
theories. And at least from the theoretical point of view, not
having a successful framework to look at `nonrelativistic'
quantum mechanics as an approximation of a `relativistic'
version of it, from the point of view of symmetry theoretical
formulation, is a shortcoming. Somewhat surprisingly, we see 
a room for the $H_{\!\ssc R}(1,3)$ symmetry to reconcile with 
most aspects of the quantum field theories. However, that put
the theories and their corresponding quantum mechanical versions 
on independent footings in conflict with the notion of the former
being the `second quantized' versions of the latter. While more 
details may have to be studied carefully, the article also 
presents some basic issues and implications about that. 

The key motivation of the study is to work towards a single
comprehensive theoretical framework that incorporates all 
experimentally established successful dynamical theories with
some as approximations of others. Results from the article, 
together with earlier efforts, score full success from the level
of `relativistic' quantum particle dynamics down to all its 
`nonrelativistic' as well as classical approximations. We 
illustrate further that the framework accommodates quantum field 
theories for high energy theory reasonably well. However, a good 
number of conceptual questions about the relationship of the 
latter  with the theories of particle dynamics remain to be 
answered. Studies in this direction also provide an approach to 
pursue the `higher level' theories beyond. Especially for that 
purpose we certainly need to further improve our understanding 
of all the theoretical and conceptual questions about the 
established theories.  
 
In the next section, we present a simple and direct illustration
of the invalidity of the on-shell mass condition so long as the 
true momentum, the canonical momentum is concerned and contrast
it with the validity for the `momentum' $m \frac{dx^\mu}{d\tau}$
as a defining property of little dynamical interest. Note that
the result is an exact contradiction of having the Poincar\'e
symmetry as the fundamental symmetry behind it because a Casimir 
invariant of the latter is violated. That is followed by 
a section on a sketch of the correct, $H_{\!\ssc R}(1,3)$, 
symmetry formulation. There are mathematically minor but important 
modifications compared to the previous presentation \cite{087}, in 
line with the parallel for the `nonrelativistic' case \cite{095}.
Details as available in Ref.\cite{087} would not be repeated here.
Discussions on how a true Poincar\'e symmetry formulation fails
and how some of the $H_{\!\ssc R}(1,3)$ symmetry features have 
actually been incorporated into the usual textbook presentations of 
the subject matter can be found. Sec.\ref{s4} then looks into the 
basic features of the Hamiltonian dynamics. The phenomenologically 
quite successful quantum field theories are `second quantized' 
versions of `relativistic' quantum mechanics, hence have related
symmetry issues. Interestingly enough, there is a way to have 
quite a reconciliation of that through the case of zero Newtonian 
mass. We present that in Sec.\ref{s5}. Some of the implications 
are among the concluding remarks presented in the last section.

\section{\boldmath Generic Canonical Momentum versus $E=mc^2$}
A practical setting to look at `relativistic' particle dynamics is the
electrodynamics of a particle of charge $e$. It is interesting that even 
the simple `nonrelativistic' Hamiltonian of
$H= \frac{1}{2m}\left(p_i- \frac{e}{c} A_i \right)\!\!\left( p^i- \frac{e}{c} A^i \right)$
actually gives the  $m \frac{dx^i}{dt}=  p^i- \frac{e}{c} A^i$,
{\em i.e.} the canonical momentum as different from mass times
velocity.  The latter is a `momentum' that has then a magnitude 
constant in time, while the same certainly does {\em not} hold for 
the true momentum vector $p^i$. From the perspective of Hamiltonian
dynamics, when  $p \ne m v$, $mv$ is not a quantity of particular
interest. We advocate against even using the word momentum for it. 
We may not then have the Newtonian form of the equation of motion 
as $F=ma$, but maintains force being the rate of change of (canonical) 
momentum and that has been generally accepted as correct. The exact 
parallel, which we prefer to write in the quantum sense, is given by
\bea
\hat{H}_s =\frac{1}{2m}\left(\hat{P}_\mu -\frac{e}{c}\hat{A}_\mu \right)\!
    \left(\hat{P}^\mu -\frac{e}{c}\hat{A}^\mu \right) \;,
\eea
is natural to be accepted, in line with the notion of $\hat{A}_\mu$ as 
a connection, to be a Lorentz covariant Hamiltonian for the `relativistic' 
case. We put an $s$ there to denote the evolution parameter, which
can be seen here as the particle proper time $\tau$. The
Hamiltonian gives
\bea
\frac{d \hat{X}^\mu}{ds}=\frac{1}{m} \left( \hat{P}^\mu -\frac{e}{c}\hat{A}^\mu \right) 
\equiv \frac{1}{m}  \hat\pi^\mu\;,
\qquad 
\frac{d}{ds} \left( \hat{P}^\mu -\frac{e}{c}\hat{A}^\mu \right)= 0\;.
\eea
Note that $\hat{P}^\mu$ being the observable in the Heisenberg
commutation relation, as essentially the Poisson bracket between 
canonical variables \cite{081}, is again the canonical momentum
and it is {not}  $m \frac{d \hat{X}^\mu}{d\tau}$, even for $s=\tau$.
The classical case is obtained simply by turning the quantum 
observables into classical ones. The equations give the correct Lorentz 
force picture. Note that $\frac{d(\pi_\mu \pi^\mu)}{ds}=0$, which is 
in line with $\pi_\mu \pi^\mu=-m^2 c^2$. The latter is, however,
a necessary consequence of the definition of $\tau$ ($=s$) as the 
time coordinate in the particle rest frame.  Hence, it has really no 
dynamical content. $\frac{d(p_\mu p^\mu)}{ds}$ cannot be zero,
and the true momentum four-vector has an evolving magnitude. 
Note that the covariant $H_s$ gives exactly the same equations
of motion as the more conventional 
$H_t= c \pi^{\ssc 0} + e A^{\ssc 0}$ with 
$\pi^{\ssc 0}=\sqrt{ \pi_i \pi^i + m^2c^2}$ on the $(x^i,p^i)$ 
phase space.

The mass in $E=mc^2$ is a notion of  `rest mass' \cite{O}.  For the 
problem at hand, $mc^2$ is the constant timelike magnitude of 
the four-vector $c \pi^\mu$ and the value of its `energy' or time
component at the particle rest frame. Yet, the charged particle is 
under an accelerating (Lorentz) force. The particle rest frame is 
hence not even an inertial frame. The quantity should not have
any particular importance, and the same goes for the proper time 
$\tau$. There is actually no good reason then to think about the 
latter as the right evolution parameter for the covariant
Hamiltonian formulation. That is the reason for us to use $s$
instead of $\tau$ to begin with. It is amazing that the simple and
clear result of not having a meaningful on-shell mass condition 
for the charged particle seems not to have received the proper 
attention and most physicists still see the condition as straightly
valid at least so long as particle dynamics is concerned.

\section{The Correct Symmetry Theoretical Formulation} 
The above result runs in direct contradiction with having 
Poincar\'e symmetry as the `relativistic' symmetry. On the other
hand, our earlier presentation of a serious group theoretical
formulation of fully Lorentz covariant quantum mechanics 
 \cite{087} was based on a different, bigger, symmetry. It is
the $H_{\!\ssc R}(1,3)$, as given by the Lie algebra
\bea &&
[J'_{\mu\nu}, J'_{\rho\sigma}] 
= i\hbar \,c \left( \eta_{\nu\sigma} J'_{\mu\rho} 
  + \eta_{\mu\rho} J'_{\nu\sigma} - \eta_{\mu\sigma} J'_{\nu\rho} 
   -\eta_{\nu\rho} J'_{\mu\sigma}\right) \;,
\sea
[J'_{\mu\nu}, Y_\rho] = i\hbar\,c   \left( \eta_{\mu\rho} Y_{\nu} 
   - \eta_{\nu\rho} Y
   _{\mu} 
  \right) \;,
  \sea
[J'_{\mu\nu}, E_\rho] = i\hbar\,c  \left( \eta_{\mu\rho} E_{\nu} 
   - \eta_{\nu\rho} E_{\mu} 
  \right) \;,
\sea
[Y_\mu, E_\nu] = i\hbar\,c  \,\eta_{\mu\nu} \,M\;,
\label{h13}
\eea
with the Minkowski four-vector indices going from $0$ to $3$, 
where $\eta_{\mu\nu}=\mbox{diag}\{-1,1,1,1\}$. That is actually
more in line with the earlier works of Refs.\cite{Z,J}. The 
presentation here has been slightly modified from that of 
Ref.\cite{087}, along the exact parallel of that of the `nonrelativistic'
case in Ref.\cite{095}. The key point is to see the central charge
generator $M$ as essentially a Casimir element, eigenvalues of
which constitutes a Casimir invariant as a label of irreducible
representations. The latter is the Newtonian mass $m$. A generator
in a Lie algebra is not usually considered a Casimir element. However, 
the central generator completely satisfies the requirement that 
an eigenspace of its operator representation would be invariant 
under any other operators of the symmetry representation, hence
giving an irreducible representation with its eigenvalue as one of
the basic characteristics. Having the Newtonian mass obtained as 
a basic characteristic of a (quantum) particle from the formulation
is a very appealing feature. In the consideration of a composite 
system, say of two particles, the formulation here further avoids 
the otherwise inescapable conclusion of having additive position 
observables.  An irreducible representation \cite{087} would 
have  $\hat{X}_\mu \equiv \frac{1}{m}\hat{Y}_\mu$ and 
$\hat{P}_\mu \equiv \frac{1}{c} \hat{E}_\mu$, satisfying the
commutation relation
$[\hat{X}_\mu, \hat{P}_\nu]= i\hbar \eta_{\mu\nu} \hat{I}$.
A composite system should correspond to the product
representation of the symmetry with all abstract generators 
$G$ of the Lie algebra represented by 
$\hat{G} = \hat{G}_a \otimes \hat{I} + \hat{I} \otimes \hat{G}_b$
where $\hat{G}_a$ and $\hat{G}_b$ are the operator 
representations of $G$ in the irreducible representations 
for particle $a$ and  particle $b$, respectively.  Hence, the
symmetry theoretical formulation gives additive $\hat{M}$,
$\hat{P}_\mu$, and $\hat{Y}_\mu$ for the corresponding
observables as physical quantities. Yet, instead of additive
$\hat{X}_\mu$, it dictates the notion of center of mass from
\bea
\hat{X}_\mu \equiv  \frac{1}{m} \hat{Y}_\mu 
= \frac{1}{m} ( m_{\ssc a} \hat{X}_{\mu a}  \otimes \hat{I} + \hat{I} \otimes m_{\ssc b} \hat{X}_{\mu b} )\;,
\eea
with $m=m_{\ssc a} + m_{\ssc b}$. The feature has been more 
or less appreciated in the Galilean symmetry picture of the
`nonrelativistic' theory, though more in the language of $U(1)$
central extension. The Heisenberg-Weyl symmetry perspective,
however, the incorrect picture of having the position observables
as directly representing symmetry generators is usually taken
(see Ref.\cite{095}). The modification is
actually complementary to the key notion under focus here.
Poincar\'e symmetry gives (Einstein rest) mass-square and
spin as Casimir invariants. The former dictates the 
non-negotiable on-shell condition for the four-vector
$\hat{P}_\mu$ of any representation. The $H_{\!\ssc R}(1,3)$
symmetry, other than giving the real consistent picture of
a formulation of Lorentz covariant quantum mechanics, 
avoids running into  a contradiction with the above result in 
electrodynamics, yet gives also a notion of particle mass. 
The latter has nothing to do with any on-shell condition, but 
agrees with the same from the `nonrelativistic' theory; both
are to be inherited in the classical theories as approximations 
to the quantum ones. 

We give a summary of the key representation theoretical results
in the appendix, focusing especially on parts beyond a simple
adaption of what has been given in Ref.\cite{087}. Here, we first 
address the matching of $H_{\!\ssc R}(1,3)$ versus the Poincar\'e
symmetry, which can be seen as its subgroup generated by 
$J'_{\mu\nu}$ and ${P}_\mu$. 

The theory of quantum mechanics from $H_{\!\ssc R}(1,3)$ can 
easily be appreciated as a natural extension of the `nonrelativistic'
one, from the analog $H_{\!\ssc R}(3)$ \cite{070,095}, with
three-vector notions extended to (Minkowski) four-vector ones. 
Note that our $H_{\!\ssc R}(3)$ symmetry is mathematically exactly 
the $U(1)$ central extension of the that of the `standard' Galilean one
with the time translation taken out. A Schr\"odinger wavefunction 
picture would have state vectors given by functions $\phi(x^\mu)$ 
with $\hat{X}_\mu$ and $\hat{P}_\mu$ acting as $x_\mu$ and 
$-i\hbar{\partial}_\mu$. The usual presentations of `relativistic' 
quantum mechanics of course share the same. But it is assumed,
rather than in any sense obtained from a symmetry theoretical
construction.  A serious symmetry theoretical construction gives
not only a vector space of states as a representation space, but 
also the matching representations of all elements of the Lie algebra, 
Lie group, universal enveloping algebra, or extensions of the group
$C^*$-algebra as operators on the vector space \cite{087,070}. 
The operators should include all (quantum) observables. The
position operators certainly cannot be obtained from the 
Poincar\'e symmetry. Assuming the usual quantum mechanical
$\hat{X}_\mu$ and $\hat{P}_\mu$ is really looking at the space
of functions $\phi(x^\mu)$ as a representation space of the $H(1,3)$ 
Heisenberg-Weyl symmetry (of ${Y}_\mu$,  ${P}_\mu$, and $M$) 
or the spin-zero representation of the full $H_{\!\ssc R}(1,3)$.
We simply cannot have the position operators from the
Poincar\'e symmetry. Notice that while we have the standard
picture of orbital angular momentum 
$\hat{L}_{\mu\nu} \equiv \hat{X}_\mu \hat{P}_\nu -\hat{P}_\mu \hat{X}_\nu$,
which is $\hat{J}_{\mu\nu}$ for the spin-zero representation,
from the $H_{\!\ssc R}(1,3)$ symmetry, there really seems to
be no way to obtained $\hat{X}_\mu$ from some operator 
representation of $\hat{P}_\mu$ and  $\hat{J}_{\mu\nu}$ 
without first putting the $\hat{X}_\mu$ in $\hat{L}_{\mu\nu}$.

One may follow closely the supposed symmetry theoretical
construction from the Poincar\'e symmetry, as for example given 
in a common textbook by Tung  \cite{T}. Starting with a (unitary) 
representation of the subgroup of translations, generated by 
$\{ P_\mu \}$, specifically a Hilbert/vector space spanned by the 
eigenvectors $\{ \left| p^\mu \rra \}$, for a particle of rest mass 
$m_{\!\ssc E} > 0$, an irreducible representation of spin-zero is 
claimed under the condition $-p_\mu p^\mu = m_{\!\ssc E}^2 c^2$
as dictated by the Casimir invariant. Here, and in the subsequent
discussions, we use $m_{\!\ssc E}$ for the Einstein rest mass to
distinguish it from the conceptually different notion of our 
Newtonian mass. Naively, the infinite-dimensional Hilbert space as 
the functional space of $\phi(x^\mu) \equiv \lla x^\mu | \phi \rra$
is identified as the representation space and such a Schr\"odinger
wavefunction is seen as related to the momentum space 
wavefunction $\tilde\phi(p^\mu) \equiv \lla p^\mu | \phi \rra$, 
through a Fourier transform from 
$\lla x^\mu | p^\mu\rra =e^{\frac{ip_\mu x^\mu}{\hbar}}$.
The truth is that many of the standard results for the so-called
position and momentum `representations' of `nonrelativistic'
quantum mechanics, as from rigorous representation theory
of the $H_{\!\ssc R}(3)$, or simply $H(3)$, symmetry have been 
implemented in a $1+3$ version.  That is, of course, effectively 
using the $H_{\!\ssc R}(1,3)$ symmetry. Otherwise, as the
Poincar\'e symmetry has no origin for the operators 
$\hat{X}_\mu$, one can have  $\hat{P}_\mu$ as $p_\mu$ but 
no $\hat{X}_\mu$ as $i\hbar \frac{\partial}{\partial p^\mu}$,
and no notion of $\left| x^\mu  \rra$ as states. In fact,
even the momentum space wavefunction $\tilde\phi(p^\mu)$,
or the Hilbert space spanned by  $\{ \left| p^\mu \rra \}$,
cannot be justified from the Poincar\'e symmetry. An
irreducible representation of the subgroup of translations has
to be one-dimensional as the subgroup, unlike the $H(1,3)$, is an 
Abelian one. Each eigenvector $\{ \left| p^\mu \rra \}$ spans one 
independent irreducible representation. Even when the Lorentz 
symmetry part is supplemented, at the best one can expect 
$\hat{J}_{\mu\nu}$ to rotate the $p^\mu$ four-vector keeping 
$-p_\mu p^\mu = m_{\!\ssc E}^2 c^2$ constant (as an irreducible
representation of $H_{\!\ssc R}(1,3)$ induced by those of its
$H(1,3)$ subgroup). That gives only 
a Hilbert space spanned by $\{ \left| p^i \rra \}$, or
$\{ \left| \vec{p} \rra \}$ as independent basis vectors, hence 
only $\tilde\phi(\vec{p})$. Then, it looks like one gets the same 
Hilbert space as that of `nonrelativistic' quantum mechanics, 
except that $\hat{X}_i$ we want still has to be imported beyond 
the Poincar\'e symmetry. Using a $\tilde\phi(p^\mu)$ notation for 
such a $\tilde\phi(\vec{p})$, assuming $-p_\mu p^\mu = m_{\!\ssc E}^2 c^2$ 
is misleading. Writing  $\phi(x^\mu)$ is much worse, because it 
is completely unclear how the $x^{\ssc 0}$ value can be obtained 
from the three independent $x^i$ values from the three-dimensional
Fourier transform. Note that the mathematics of the latter cannot 
automatically be taken to give physical meaning to the $x^i$ 
variables either. And in the actual physical applications, one 
'naturally' thinks about having the full $x^\mu$ in $\phi(x^\mu)$ 
as a free Minkowski four-vector, which certainly cannot be obtained 
from the construction, though Minkowski spacetime can indeed be 
obtained from the Poincar\'e symmetry as a coset space representation. 
Taking the theory as one of, representation of, the $H_{\!\ssc R}(1,3)$ 
symmetry obviously makes better sense. As none of the $\hat{X}_\mu$ 
commutes with the  $\hat{P}_\mu\hat{P}^\mu$, an $\hat{X}_i$ action 
on the quantum state of a particle would take the state beyond an
eigenspace of $\hat{P}_\mu\hat{P}^\mu$ hence violating the on-shell
mass condition and pulling the different eigenspaces from different
irreducible representations of the Poincar\'e symmetry (as 
a subgroup) into a single (induced) irreducible representation 
of $H_{\!\ssc R}(1,3)$. A Hamiltonian evolution with an 
$\hat{X} _\mu$-dependent Hamiltonian is just an example of that. 
That is really saying, a notion of a position observable cannot
make sense for a particle with the Einstein rest mass as a fixed
characteristic as the corresponding operator cannot be defined
within its vector space of states. Without the position observables
and the Heisenberg commutation relation between them and the
$\hat{P}_i$, one has hardly a particle theory, nor a quantum one.

\section{A Look at the Covariant Hamiltonian Dynamics \label{s4}}
A covariant Hamiltonian formulation of classical and quantum
dynamics has been taken by some physicists starting probably
from the works of Fock and Str\"uckelberg in the 40's. For
a general picture of the lines of work, and sources of further
references, books \cite{TS,Ho,F} may be consulted. Most authors,
however, has struggled to maintain the on-shell mass condition 
for the momentum four-vector in some way. On the other hand,
a no-interaction theorem has been derived from imposing the
straight on-shell mass condition for the particle involved \cite{ni}.
The latter is unavoidable when each particle is seen as from an
irreducible representation of the Poincar\'e symmetry. Despite
that, quantum field theories supposed to be obtained from the
symmetry talk about `virtual' states which violate the condition.
From the perspective here, all such theories should be re-examined 
from an $H_{\!\ssc R}(1,3)$ symmetry point of view.

A truly covariant Hamiltonian formulation should have free
four-vector canonical variables, the quantum $\hat{X}^\mu$ and
$\hat{P}^\mu$ and the ${x}^\mu$ and ${p}^\mu$ in the classical
approximation. A typical single-particle Hamiltonian would have 
the form  $\hat{H}_s =\frac{\hat{P}_\mu \hat{P}^\mu}{2m} + V(\hat{X}^\mu)$
which dictates Hamiltonian flows in the phase space each as 
a one-parameter group of transformations with an evolution
parameter $s$. The latter is mathematically the exact analog 
of the Newtonian time in the `nonrelativistic' case so long as 
the Hamiltonian formulation is concerned \cite{095}, though 
it is not clear if it is some sort of physical time at all. One has 
$\frac{d}{ds}= -\frac{1}{i\hbar}[ \hat{H}_s, \cdot]$, the 
Hamiltonian vector field in terms of the Poisson bracket as
in the classical case \cite{081}. (In fact, $\hat{X}^\mu$ and
$\hat{P}^\mu$ can be seen as noncommutative coordinates
of the exact phase space of the quantum states in the usual
Schr\"odinger picture \cite{081,087,078}). For the free
particle case, with ${\hat{P}_\mu \hat{P}^\mu} = -m_{\!\ssc E}^2 c^2$
here as the fixed initial value, we have $\frac{d \hat{P}^\mu}{ds}=0$
and $\frac{d \hat{X}^\mu}{ds}=\frac{1}{m} \hat{P}^\mu$,
at least easy to interpret for initial momentum eigenstates or 
in the classical limit. Note that  the operator $d \hat{X}^\mu$
does not commute with  $\hat{X}^\mu$, from the perspective
of the noncommutative differential geometry \cite{078} or
otherwise. We have a constant $m_{\!\ssc E}$ actually 
satisfying $\frac{d\tau}{ds}=  \frac{m_{\!\ssc E}}{m}$. Identifying 
the Einstein rest mass $m_{\!\ssc E}$ with  the Newtonian
mass $m$ gives $s$ as $\tau$, in agreement with the usual
textbook descriptions (more of the classical case). Note that the 
$s$-independent Schr\"odinger equation (for a spin-zero free
particle) gives exactly the form of the Klein-Gordon equation as
$\hbar^2 \partial_\mu \partial^\mu \phi(x^\mu) ={m_{\!\ssc E}^2 c^2} \phi(x^\mu)$.

With the $H_{\!\ssc R}(1,3)$ symmetry, as well as its classical
approximation, $P_\mu$ generate translations in $x^\mu$ and
$Y_\mu$ generate translations in $p^\mu$. Such translational
symmetries can be taken as coordinate transformations on the 
phase spaces. As such, they are really reference frame 
transformations. In fact, from the perspective of quantum reference 
frame transformations \cite{qrf,093}, or that of our phase space
as quantum Minkowski spacetime \cite{087,078}, it is especially
clear that the notion of reference frames in a dynamical theory
should really be one for the phase space. Of course, the Lorentz
transformations are also valid reference frame transformations.
With that understanding, one can see that taking any value of
$m_{\!\ssc E}$ to the Newtonian mass $m$ can be achieved with
a  reference frame transformation. Let us focus on the classical 
case as it is more straightforward to interpret.  For example, in 
a particle rest frame, a translation in $p^{\ssc 0}$, or the energy
value, is exactly one in the $m_{\!\ssc E}$ value. We are familiar 
with an arbitrary absolute value of energy at least in Newtonian 
physics. It is easy to appreciate that we can measure in practice 
only energy exchanges but never an absolute energy value. In fact, 
without being able to do that, there is hardly any way to check the 
exact validity of the on-shell mass condition experimentally. Here, 
the symmetry picture suggests seeing setting a reference to 
define the absolute energy value as a choice of frame of reference 
for the Lorentz covariant phase space. 

In a (classical) situation with a nontrivial interaction, with 
a nonzero potential $V(x^\mu)$ or under an electromagnetic
field, the $m_{\!\ssc E}$ or $p_\mu p^\mu$ cannot be kept 
constant. It goes without saying that any fixed choice of reference 
frame can only put $m_{\!\ssc E}$ as $m$ for one value of $s$,
maybe only for one of the particles involved.
An interesting analysis of two interacting particles has been 
presented in Ref.\cite{PR}. The Hamiltonian is given as
\bea
\frac{p_{a\mu}p_a^\mu}{2m_a} + \frac{p_{b\mu}p_b^\mu}{2m_b}  
     +V(|x_a^\mu - x_b^\mu|)
= \frac{p_{\mu}p^\mu}{2m} + \frac{q_{\mu}q^\mu}{2\mu}  
     +V(r_\mu r^\mu)  \;,
\eea
where we have $p^\mu=p_a^\mu +  p_b^\mu$ as the conserved 
total momentum, canonical variable conjugated to the center of 
mass position $x^\mu= \frac{m_a x_a^\mu +  m_b x_b^\mu}{m}$, 
$m=m_a + m_b$, and $r^\mu$-$q^\mu$ being the other pair of
canonical variables, for the relative motion degree of freedom,
given by $r^\mu = x_a^\mu - x_b^\mu$, and 
$q^\mu= \frac{m_a p_b^\mu -  m_b p_a^\mu}{m}$, with 
$\mu= \frac{m_a  m_b}{m}$ being the reduced mass. Actually, 
for the two particles with an initial spacetime separation that is 
spacelike, {\em i.e.} with $r_\mu r^\mu >0$, we can use a Lorentz 
transformation and a momentum, or rather energy, translation 
symmetry to choose a frame of reference putting the initial values 
of $r^{\ssc 0}$ and $q^{\ssc 0}$ to zero, so long as the two particles
do not have exactly the same mass. Otherwise, as the authors 
noted, when the condition $j_{\mu\nu}j^{\mu\nu}>0$, with
$j^{\mu\nu} \equiv r^\mu q^\nu - r^\nu q^\mu$,  is satisfied one 
can achieve the same with simply a Lorentz boost. Hamilton's 
equations of motion then say $r^{\ssc 0}$ and $q^{\ssc 0}$ stay 
zero, hence the nontrivial equations of motion, for the relative 
degree of freedom, are really just
\bea
q^i= \mu \frac{dr^i}{ds} \;,
\qquad
 \frac{dq_i}{ds} = - \frac{\partial V(r_i r^i)}{\partial r^i} \;,
\eea
and $s$ the evolution parameter for the Hamiltonian.
We can have then
$x_a^{\ssc 0}=x_b^{\ssc 0}= x^{\ssc 0}= \frac{p^{\ssc 0}}{m} s$,
where $p^{\ssc 0} c$ is the conserved total energy for the system,
giving $s$ as a rescaled value of the common coordinate time 
for the particles. The dynamics is then essentially the same as 
the Newtonian theory. Such frames may not be compatible with 
$p_{\mu}p^\mu = -m^2c^2$ though. Note that what we used to
think about as `relativistic' dynamics is essentially only about  
spacelike initial spacetime separation. More careful analyses 
of other practical dynamical situations with the new symmetry 
perspective taken into consideration are certainly of interest.

The basic picture of covariant Hamiltonian formulation, to the 
extent of the actual Hamiltonians discussed, we have made no
new contribution. That has all been available in the literature.
For the general validity of the theories, classical or quantum,
beyond the few cases discussed explicitly here, it suffices to say
that there is no problem at all for the formulation to at least
incorporate the successes of the usual formulation. At the classical
level, the textbook by Johns \cite{Jh}, for example,  discussed 
in some detail how the usual Hamiltonian formulation can be
cast in the 'equivalent' alternative of the covariant formulation
in the 'extended' phase space. In the quantum case, the 
$s$-independent Schr\"odinger equations of motion can be 
seen as exactly the usual equations. As discussed above, the 
$\phi(x^\mu)$ wavefunction though can only truly be justified 
from our symmetry perspectives presented. The current paper
focuses on the symmetry theoretical and background conceptual
issues rather than the practical ones of applications.  Otherwise,
the dynamical theory certainly can describe physics beyond 
what can be done in the framework of the usual formulation. 
Some works in this direction are already available in the 
literature, though in our opinion, not enough attention has
been paid on clarifying the physics pictures better. The current
article aims exactly at providing a better conceptual background
for such a task.

\section{\boldmath Quantum Field Theories from  $H_{\!\ssc R}(1,3)$ Symmetry \label{s5}}
Quantum field theories are formulated based on the perspective of 
`second quantization', promoting the notion of a wavefunction as the 
description of a quantum multi-particle state to an operator that   
annihilates it, and the Hermitian conjugate of the wavefunction to an 
operator that creates the state from the vacuum as the ground state. 
So, a quantum field theory for a quantum field is the `quantization' 
of a `first quantized', or `quantum particle', theory. From the point 
of view of symmetry theoretical formulation, the `single-particle'
state wavefunction should correspond to an irreducible representation
of the symmetry and the vector space of states, usually to be taken 
as a Hilbert space, is the representation space. It is exactly in this 
sense that textbooks on quantum field theories talk about the quantum 
fields as irreducible representations of the Poincar\'e symmetry. The 
`multi-particle' states for a quantum field at a fixed value of the 
`number of particles' corresponds to an invariant part, with the proper 
exchange symmetry, of a product representation. However, besides our
critique about the `first quantized' `relativistic' particle theory,
for gauge fields such as the electromagnetic field, the straight logic
of `second quantization' is at odd with the classical picture of fields. 
There is certainly no two-step quantization for the latter. Whether the 
`first quantized' electromagnetic field $A_\mu(x^\nu)$ is a classical 
field or a quantum wavefunction of the photon as a `particle', or if 
some quantum wavefunctions can be seen as classical fields, seems 
never been properly addressed. `Massless' classical particles are 
certainly not admissible in `nonrelativistic' theories. It is not 
clear either if there can be sensible theories about them without 
invoking quantum notions even in the `relativistic' setting. 

In general, a quantum field theory cannot be seen as corresponding 
to a representation of the background (relativity) symmetry. The 
creation and annihilation operators, unlike the single- or 
multi-particle operators, cannot be obtained in any way from a
representation picture of that symmetry. For a symmetry picture,
they, or at least Hermitian linear combinations of them, would 
be like generators of extra symmetries which extend the background
(relativity) symmetry of the corresponding `first-quantized' theory 
to a bigger symmetry. The quantum field theory would an irreducible 
representation of the extended symmetry. Without a picture of the 
latter, we actually do not have full symmetry theoretical formulation 
for a quantum field theory, not to say a high energy theory involving 
a list of quantum fields with nontrivial interaction among them. 
Though the `particle' terminology is still in common use in quantum 
field theories, `multi-particle' states certainly have very different 
physics beyond particle dynamics. At least from the symmetry 
theoretical formulation point of view, the Newtonian notion of 
a particle as an elementary constituent of the physical theory and 
a point mass cannot be applied. Moreover, at least in principle, the 
state space of a quantum field theory contains states which are 
nontrivial linear combinations of `multi-particle' states of 
a different `number of particles'. Note that direct experimental 
results from high energy physics are essentially all obtained at
the classical limit.
 
While it may be interesting to construct a new quantum field theory
from the new theories of `relativistic' quantum mechanics discussed,
we do not intend to do that here. The existing quantum field theories
seem to have very successful applications in high energy physics of 
the Standard Model. We want to see if we can have some reconciliation 
of that with the new relativity symmetry picture of $H_{\!\ssc R}(1,3)$. 
With the above clarification of the, actually quite limited, relation 
of a practical quantum field theory in high energy physics and the 
background symmetry, we are ready to illustrate how that can work.
To focus more on the key features, working in the simpler setting 
of the real scalar field serves the purpose best. 

Representation theory of the $H_{\!\ssc R}(1,3)$ has a special, kind 
of degenerated, case. That is the case of zero Newtonian mass (as the
Casimir invariant from generator $M$). The case is tricky. For the key
part of $H(1,3)$, it is usually given as representation for the Lie 
group \cite{T}. But at the Lie algebra level, it is like trivializing 
the Lie brackets, similar to the contraction to the classical limit 
or removing the relevance of the central extension. Moreover, as the 
position observables are to be introduced on the representation space 
as $\frac{1}{m}\hat{Y_\mu}$, they cannot be properly defined. Yet, 
when we are looking at fields instead of particles, that is not 
a problem. While the notion of a field has a background picture of 
spacetime, it should not be expected to have a specific position in 
it. For each value of $m \ne 0$, we have an irreducible representation. 
For $m=0$, we have actually one-dimensional irreducible representations 
for the $H(1,3)$ group that has a group element 
$e^{\frac{i}{\hbar} ( \tilde{v}^\mu Y_\mu - \tilde{x}^\mu P_\mu + \tilde{u} M)}$
represented by the operator $e^{\frac{i}{\hbar} ( \tilde{v}^\mu y_\mu - \tilde{x}^\mu p_\mu )}$,
for real four-vector parameters $\tilde{v}^\mu$ and $\tilde{x}^\mu$
 \cite{T}. It is easy to appreciate. When $\hat{M}$ is effectively 
zero, all operators representing elements of the Lie algebra commute. 
They can be taken as real numbers acting, as scalar multiplications, 
in a one-dimensional vector space to make the representations unitary. 
Consider such representation spaces as the span of a single $\hat{P}_\mu$ 
eigenstate, the action of a $\hat{Y_\mu}$ as multiplication of a real 
number $y_\mu$ need not be much of a concern for moving towards the 
`second quantization' framework. Extending an irreducible representation 
of $H(1,3)$ to a spin-zero representation of the $H_{\!\ssc R}(1,3)$,
different representation spaces of $\left| p_\mu \rra$ related by 
Lorentz transformations are pulled together as a single irreducible 
representation. The resulting representation space is the space of 
$\left| p_\mu \rra$ with constant Lorentz invariant
 $-m_{\!\ssc E}^2c^2 = p_\mu p^\mu$, or rather  the space of $\left| \vec{p} \rra$. 
It is exactly like what we discussed above for the Poincar\'e symmetry
case. In fact,  $\hat{P}_\mu \hat{P}^\mu$ now commutating with all
operators representing elements of the $H_{\!\ssc R}(1,3)$ Lie algebra 
serves as an Casimir operator for the representation. 

The $m=0$ irreducible representations of $H_{\!\ssc R}(1,3)$ seem
to have little structure even in comparison to a picture of classical
mechanics. It is hard to think about the systems described as quantum.
Yet, that may be enough for implementing the so-called `second 
quantization', which may proceed as follows. Take a $\left| \vec{p} \rra$
state as the `single-particle' state. Introduce the creation and
annihilation operators $\hat{a}_{\vec{p}}^\dag$ and $\hat{a}_{\vec{p}}$
and the vacuum state $\left|0\rra$ with
\bea
\hat{a}_{\vec{p}} \left|0\rra =0 \;,
\qquad
\left| \vec{p}_1 \dots \vec{p}_n  \rra = \sqrt{2E_{\vec{p}_1}\dots 2E_{\vec{p}_n}}
   \hat{a}_{\vec{p}_1}^\dag \dots \hat{a}_{\vec{p}_n}^\dag \dots \left|0\rra \;,
\eea
where $E_{\vec{p}_n}\equiv \sqrt{|\vec{p}|^2 + m_{ E}^2c^2} (c)$,
together with the Lorentz invariant inner product
$\lla \vec{p}_1 | \vec{p}_2 \rra= {2E_{\vec{p}_1}} (2\pi)^3 \delta^3(\vec{p}_1-\vec{p}_2)$.
Of course, the usual commutation relations among those creation and
annihilation operators are also assumed. What one has then is 
essentially a picture of quantum field theory, as for example presented 
in Maggiore \cite{M}. In fact, it is exactly this momentum eigenstate 
picture that is mostly used in applications like Feynman diagram 
calculations. One can even define the field operator
$\phi(x) \equiv \int\!\!\frac{d^3p}{(2\pi)^3 \sqrt{2E_{\vec{p}}}}
  ( e^{-ipx} \hat{a}_{\vec{p}} + e^{ipx} \hat{a}_{\vec{p}}^\dag )$,
as a linear combination. The basic mathematical framework of the usual 
quantum field theory is thus obtained, though under a quite different 
line of thinking. 

We have seen above that the so-called `second quantization' offers a
procedure to construct the `quantum field theories' that requires very
little background. A usual picture of some `first quantized' quantum
particle theories may be by-passed from the perspective of a symmetry
theoretical formulation. It can work from the $m=0$ irreducible 
representations of $H_{\!\ssc R}(1,3)$ which in themselves hardly give 
an interesting dynamical setting, unlikely the $m\ne 0$ cases we 
discussed in the previous section. 


\section{Concluding Remarks} 
It should be clear that in all considerations of `relativistic' 
quantum mechanics, one always has the position and momentum 
observables satisfying the Heisenberg commutation relation incorporated, 
though probably only the spatial part may be taken explicitly. That is, 
of course, what we need in the theories but not what a first principle 
Poincar\'e symmetry theoretical formulation can offer.  We are really 
dealing with representations of $H_{\!\ssc R}(1,3)$ but in the name 
of its subgroup of Poincar\'e symmetry. The magnitude (square) of the 
momentum four-vector changes under the action of a position operator 
and hence cannot be a constant of the theory. Physicists keep thinking 
the latter is the right symmetry, we think, mostly for two main reasons. 
First of all, the symmetry seems to be the natural symmetry of Minkowski 
spacetime and seems to be fine for the classical mechanics though one 
cannot quite have a covariant phase space (see Ref.\cite{086}). 
Secondly, it is mostly from the on-shell mass condition, though really
that of free particles, that we appreciate the famous equation $E=mc^2$
giving the great understanding that mass is a form of energy. Poincar\'e
symmetry dictates the condition. $H_{\!\ssc R}(1,3)$ actually does not 
admit that. And naturally, most physicists do not check through the 
straight logic of a rigorous symmetry theoretical formulation, yet easily 
accept a claim that it works. From our presentations, it should be clear 
that it is the $H_{\!\ssc R}(1,3)$ that really is the right symmetry. We 
have discussed the perspective that it is the symmetry for the particle 
phase space that gives a full formulation of a theory of particle dynamics
that should be taken as the true relativity symmetry of the theories
 \cite{095}. A picture of the space(time) as some may naively conceive, 
other than the picture of the configuration (sub)space of 
single-particle phase space is not physical and of no interest to 
physics so long as particle dynamics is concerned. While logically
speaking, the notion of spacetime in a field theory may be different,
how one can justify a physical notion of spacetime based completely on 
the mathematical formulation of a field theory in itself looks, in our
opinion, quite problematic. We have presented earlier a consistent 
theory of Lorentz covariant quantum mechanics from the symmetry 
theoretical formulation based on $H_{\!\ssc R}(1,3)$ \cite{087}. At the 
proper symmetry contraction limits, the full dynamical theory gives the
`nonrelativistic' and the classical theories as approximations. We 
further highlighted in this article that the on-shell mass condition 
actually does not hold even in the simple case of the dynamics of a 
charged particle under a background electromagnetic field, and is 
violated by the action of any position operator. The case for 
$H_{\!\ssc R}(1,3)$ against the Poincar\'e symmetry should be 
seen as firmly established.

Particle dynamics from the $H_{\!\ssc R}(1,3)$ and its classical
limit gives no constant Einstein rest mass in general. Though one
can choose a reference frame to put any initial value of it to the 
fixed Newtonian mass $m$, it may not be maintained in the dynamical 
evolution. That can clearly be established from the dynamics of 
a charged particle under a Lorentz force. The key is to identify 
the correct momentum as the canonical momentum. Mass as a form of 
energy is a great insight from Einstein, which we do not contest. 
But there is no fixed value of the magnitude of the momentum 
four-vector given to be the particle mass or otherwise. 
$m \frac{dx^\mu}{d\tau}$ has constant magnitude by definition. 
But that has no dynamic content. The actual evolution parameter 
$s$ from a general covariant Hamiltonian formulation may have 
a nontrivial relation to any particle proper time $\tau$, 
{\em i.e.} $\frac{d\tau}{ds}$ is not necessarily a constant in 
a generic situation. The latter depends on the actual interaction 
potential, which is not allowed from a Poincar\'e symmetry 
formulation with the on-shell mass condition. Up to this point, 
we do not have a practical situation where such a particle 
interaction picture applies. However, we do have solid 
applications of the kind of particle interaction picture in 
Newtonian mechanics, and the quantum analog. The corresponding 
particle interaction picture from the $H_{\!\ssc R}(1,3)$ 
formulation seems to be the only Lorentz covariant quantum 
particle dynamics picture from which the `nonrelativistic' 
theories can be retrieved as approximations.

The dynamical `velocity' $\frac{dx^\mu}{ds}$ is not the Einstein 
four-velocity $\frac{dx^\mu}{d\tau}$. We want to emphasize that, from 
the Hamiltonian formulation, a choice of Hamiltonian $H_s$ comes along 
with its evolution parameter $s$ which is non-negotiable. In a dynamical 
setting involving multiple particles and even fields, for example, we 
do not have to guess at the relation between $s$ and the other physical 
time parameters such as the various individual particle proper times. 
Nor do we have to ponder the question if the latter can be correlated 
{\em a priori}. Find the correct $H_s$ and the rest follow from the 
mathematics. The correct $H_s$ can be found from practical considerations
or, on the theoretical side, from feasible generalizations of successful 
`nonrelativistic' theories and checked from the validity of the results.
Any mathematical $s$-evolution results may have to be physically 
interpreted in terms of some time evolution picture, under a choice 
of frame of reference (for the phase space). 

A very practical `relativistic' theory is the theory of electrodynamics.
However, a consistent formulation of the classical theory in the general
setting of more than one charged particle is quite a nontrivial matter
 \cite{R}. There is also a version of it without the electromagnetic 
field as a true independent constituent \cite{FW}. Otherwise,
electrodynamics generally involved interacting particles and the 
electromagnetic field. Our current theory of quantum electrodynamics
is one of interacting quantum electromagnetic (or photon field) and 
quantum fields of `charged particles' like the electron. We illustrate 
above how quantum field theories may be obtained from applying some 
form of `second quantization' to $m=0$ irreducible representations 
of $H_{\!\ssc R}(1,3)$. Yet, the latter hardly looks like any `first 
quantized' particle theory. Actually, from the perspective of the usual 
formulation as given in textbooks, so long as a classical field theory 
is concerned, there is really no two-step quantization. That picture
of `quantization' of the classical electromagnetic field and the 
`second quantization' leading to the matter fields like the electron
field are conceptually quite different. We do not really have a 
theoretically and mathematically rigorous way to formulate our theory 
of quantum electrodynamics as the quantization of (one of the) theory 
of classical electrodynamics, or the latter as the classical 
approximation of the former. Other quantum fields such as the quark
fields and the nonabelian gauge fields have no practical classical
picture anyway. A comprehensive formulation framework for all the
fundamental theories from Newtonian mechanics to Standard Model high
energy theory is still a task to be accomplished, not to say with
quantum gravity included. Our current study is only a small effort 
in relation.

A comment on the two different notions of masses is in order. While 
it is clear that we have given a nice symmetry theoretical notion 
for the Newtonian mass $m$ in both `nonrelativistic' \cite{095} and 
`relativistic' quantum setting, as a very different notion from the 
Einstein rest mass $m_{\!\ssc E}$. For a quantum field theory as 
discussed, we have $m=0$ and an independent $m_{\!\ssc E}$ as 
characteristic properties. One may worry about their compatibility.
For this matter, it is interesting to note that  all the quantum 
fields other than the Higgs doublet in the Standard Model of high 
energy physics are actually massless. They are chiral fermionic 
fields and gauge fields nonzero masses for which are ruled out by
the gauge symmetries. The Higgs doublet has a negative mass-square.
The masses in quantum field theories are really energy, or 
renormalization, scale-dependent parameters. The only possible 
constant value as basic characteristic properties of the entities
is the kind of symmetry-protected vanishing value. The negative 
mass-square parameter for the Higgs doublet is never a notion of
physical characteristic. Besides, the doublet is widely believed 
to be a composite. All physical masses come from interaction 
couplings to the doublet after spontaneous electroweak symmetry
breaking. So, apart from the general fact that a more serious and
mathematically rigorous understanding of the relationship among the
various theories, from the point of view of the symmetry theoretical 
formulation or otherwise, still has to be studied, the analysis
presented in the article does not post new compatibility issues 
for `mass'.

We have mentioned above the admissible noncommutative geometric picture 
the phase space for quantum mechanics with the position and momentum 
operators as noncommutative coordinate observables \cite{087,078}. The 
perspective, when taken seriously, certainly asks for different thinking 
about the existing field theories. The Schr\"odinger wavefunction is 
really a set of complex number coordinates for the phase space, for the 
description of a state in quantum mechanics as a point inside. In our
opinion, there has never been any good reason to think about the real
`position' variables for the wavefunction as coordinates for the physical
space other than as some form of the classical approximation of such. 
The position operators as observables describing the particle position
should rather be the natural choice. The idea of quantum fields being 
fields of the classical, Minkowski, model of spacetime may be much 
restricted to looking at the quantum theory from a far too classical 
perspectives. We do not look into the construction of quantum field 
theory as a `second quantized' version of quantum mechanics from the 
$H_{\!\ssc R}(1,3)$ symmetry perspective and the noncommutative 
geometric picture. That is because we see up to this point no good reason 
for believing in the direction. After all, our main aim here is to look 
at the proper symmetry theoretical formulation of `relativistic' quantum 
physics and its implications. `Second quantization' in itself, as 
illustrated, is not much of a symmetry theoretically based notion.

We have brought up many questions about consistent conceptual and 
theoretical pictures of Lorentz covariant theories of quantum physics
fully compatible with symmetry theoretical perspectives. A lot more
effects are certainly needed to clarify that all. The existing quantum
field theory is never short of conceptual difficulties. Many have taken 
the practical approach of Feynman not to worry about that. However, to 
push for further advance in fundamental theoretical physics, it is 
important to take up the kind of question more seriously.

\section*{Appendix: Supplement to the symmetry theoretical formulation: particles with nonzero spin. }
Here, we sketch some more results from the $H_{\!\ssc R}(1,3)$
symmetry theoretical formulation, along the line of \cite{095},
including explicitly irreducible representation for nontrivial spin cases.  

We start with an irreducible representation of the 
Heisenberg-Weyl subgroup $H(1,3)$, based on (pseudo-)Hermitian 
operators $\hat{X}_\mu \equiv \frac{1}{m}\hat{Y}_\mu$ and 
$\hat{P}_\mu \equiv \frac{1}{c} \hat{E}_\mu$, with 
$[\hat{X}_\mu, \hat{P}_\nu]= i\hbar \eta_{\mu\nu} \hat{I}$ 
for $m$ being the eigenvalue of the effectively Casimir operator 
$\hat{M}$. The representation is essentially as the one presented 
in Ref.\cite{087}, with only necessary easy to trace adjustments 
involving the identification of $M$ as the central charge generator.
Similar to the $H_{\!\ssc R}(3)$ case, an irreducible representation
of $H_{\!\ssc R}(1,3)$ is given by as a direct product of one for the 
Heisenberg-Weyl group $H(1,3)$ and one for an $SO(1,3)$, or its 
double cover $SL(2,C)$, generated by
$T'_{\mu\nu} \equiv M J'_{\mu\nu} -   ({Y}_\mu {E}_\nu - {Y}_\nu {E}_\mu)$.
We have $\hat{S}_{\mu\nu}= \frac{1}{mc} \hat{T}'_{\mu\nu} =\hat{J}_{\mu\nu} - \hat{L}_{\mu\nu}$,
where $\hat{J}_{\mu\nu}= \frac{1}{c} \hat{J}'_{\mu\nu}$ and
$\hat{L}_{\mu\nu} \equiv \hat{X}_\mu \hat{P}_\nu -\hat{P}_\mu \hat{X}_\nu$.
The corresponding two Casimir operators, 
$\frac{1}{2}\hat{S}_{\mu\nu} \hat{S}^{\mu\nu}$
and $\frac{1}{4} \epsilon^{\mu\nu\rho\sigma} \hat{S}_{\mu\nu} \hat{S}_{\rho\sigma}$ 
have eigenvalues 
$2\hbar^2 [s_{\!\ssc L}(s_{\!\ssc L}+1)+s_{\!\ssc R}(s_{\!\ssc R}+1)]$
and $2\hbar^2 [s_{\!\ssc L}(s_{\!\ssc L}+1)-s_{\!\ssc R}(s_{\!\ssc R}+1)]$,
respectively, for the labeling with the left-handed and 
right-handed spins $\{m,s_{\!\ssc L},s_{\!\ssc R}\}$ of 
dimension $(2s_{\!\ssc L}+1)(2s_{\!\ssc R}+1)$. Explicitly, we have 
$\hat{S}_{ij}^{\pm} = \frac{1}{2} (\hat{S}_{ij}\pm i{\epsilon_{ij}}^k\hat{S}_{{\ssc 0}k})$,
from ${T'}_{ij}^{\pm} = \frac{1}{2} (T'_{ij}\pm i {\epsilon_{ij}}^{k} T'_{{\ssc 0}k})$,
(here ${\epsilon_{ij}}^{k}= \epsilon_{ijh}\delta^{hk}$, with 
$\epsilon_{ijh}$ being the totally antisymmetric tensor),
as generators of the two independent $SO(3)$ under the
isomorphism $SO(1,3) \simeq SO(3) \oplus SO(3)$ as complex 
Lie algebra. The picture of irreducible representations for
particles with nontrivial spins having the same form of 
wavefunctions as a zero-spin wavefunction $\phi(x^\mu)$
for each spin component though familiar from textbooks
cannot be fully justified from Poincar\'e symmetry picture
as the group cannot be seen as a direct product. Without the
generators $Y_\mu$ in the Lie algebra, the $T'_{\mu\nu}$
do not exist as elements in the universal enveloping algebra.

To get to the approximation with Lorentz boosts replaced by 
Galilean ones, we take the Lie algebra and the representation 
of interest to the $c \to \infty$ limit of 
$K_i= \frac{1}{c^2} J'_{i{\ssc 0}}$,  $P_i = \frac{1}{c} E_i$,  
$U =\frac{-1}{c} Y_{\ssc 0}$, and $J_{ij}= \frac{1}{c} J'_{ij}$
with the renaming $H\equiv -E_{\ssc 0}$.  Mathematically, 
it is to be formulated as a symmetry contraction \cite{087},
taken from the Lie algebra to the group, the proper extension
of the universal enveloping algebra or group $C^*$-algebra 
and their relevant representation as a single representation
 \cite{070,087} for the full dynamical theory, including the
symplectic structure. In this case, it is easy to see that the 
generator set $\{ J_{ij}, K_i, P_i, H \}$ gives exactly a Galilean
group $G(3)$ without the central extension. The  set 
$\{ J_{ij}, Y_i, P_i, M \}$ gives an $H_{\!\ssc R}(3)$ as the 
symmetry for the Galilean quantum particles.  We have
\bea 
[U , H]= -i\hbar  M \;,
\quad\quad
[K_i, Y_j]=  -i\hbar \delta_{ij}  U \;,
\eea
as the extra nontrivial Lie products among the full set. 
For the irreducible representation  $\{m,s_{\!\ssc L},s_{\!\ssc R}\}$,
we have $\hat{P}_i$ directly representing $P_i$, $\hat{H} =- c \hat{P}_0$ 
which apparently goes to infinite limit, and 
$\hat{T} = \frac{1}{m} \hat{U}  =\frac{-1}{c} \hat{X}_{\ssc 0}$
which goes to the vanishing limit. 
$\hat{X}_i$ and $\hat{S}_{ij}$ are unchanged. 
$\hat{J}_{ij} = (\hat{X}_i \hat{P}_j - \hat{P}_i \hat{X}_j) + \hat{S}_{ij}$
is retrieved. $\hat{K}_i=\frac{1}{c} \hat{S}_{i{\ssc 0}} 
   -\frac{1}{c^2}\hat{X}_i \hat{H} + \hat{P}_i \hat{T}$
which has the limit as $\hat{P}_i \hat{T}$. If one put the
parallel Lorentz boost to Galilean boost contraction onto the
Lie algebra of the $\hat{S}_{\mu\nu}$, one can easily see that
the left- and right-handed spin generators both become 
essentially just $\hat{S}_{ij}$ (with a factor $\frac{1}{2}$,
which makes no practical difference). Of course the particle 
mass as the key Casimir invariant maintains. Hence, one obtains 
$\{m,s\}$ of  $H_{\!\ssc R}(3)$ from $\{m,s_{\!\ssc L},s_{\!\ssc R}\}$. 

For a composite system, parallel analysis as shown above for the
$H_{\!\ssc R}(3)$ symmetric `nonrelativistic' case obviously goes
 through, with additive mass $m=m_a+m_b$, the notion of center 
of mass and $\frac{1}{2}\hat{S}_{\mu\nu} \hat{S}^{\mu\nu}$, 
here together with and also 
$\frac{1}{4} \epsilon^{\mu\nu\rho\sigma} \hat{S}_{\mu\nu} \hat{S}_{\rho\sigma}$, 
as effective Casimir operators for the composite are obtained  
from the product representation. $\hat{S}_{\mu\nu}$ for the 
contains the `orbital' angular momentum 
$\hat{R}_\mu \hat{Q}_\nu - \hat{Q}_\mu \hat{R}_\nu$, which
is a contribution to the internal or intrinsic angular momentum 
for the composite. $\hat{R}_\mu$ and  $\hat{Q}_\mu$ are 
position and momentum variables for the relative motion
degree of freedom, respectively. We have
$\hat{S}_{\mu\nu}= \hat{S}_{a\mu\nu}+\hat{S}_{b\mu\nu}
+\hat{R}_\mu \hat{Q}_\nu - \hat{Q}_\mu \hat{R}_\nu$
and $\hat{J}_{\mu\nu}=\hat{X}_\mu \hat{P}_\nu - \hat{P}_\mu \hat{X}_\nu$
with $\hat{X}_\mu$ and  $\hat{P}_\mu$ being the center of
mass position and momentum.

\bigskip
\noindent
 {\em Acknowledgments: \ \ }
The work is partially supported by research grants 
number   110-2112-M-008-016
and  111-2112-M-008-029
of the MOST of Taiwan. OK also has the support of a Visiting Professorship
at Korea Institute for Advanced Study during the later phrase of the study.


\begin{thebibliography}{99}
\bibitem{O}
L.B. Okun, Mass versus relativistic and rest masses,
Am. J. Phys. {\bf 77} (2009) 430-431.
\bibitem{H}
E. Hecht, How Einstein confirmed $E_0=mc^2$,
Am. J. Phys. {\bf 79} (2011) 591-600.
\bibitem{E}
A. Einstein, Equivalence of mass and energy,
Bull. Am. Math. Soc. {April 1935} (1935) 223-230
\bibitem{070}
C.S. Chew, O.C.W. Kong, and J. Payne,
Observables and Dynamics Quantum to Classical from a Relativity Symmetry and Noncommutative-Geometric Perspective, 
J. High Energy Phys. Gravit. Cosmol. {\bf 5} (2019) 553586.
\bibitem{087}
S. Bedi\'c,   O.C.W. Kong, and H.K. Ting, 
Group Theoretical Approach to Pseudo-Hermitian Quantum
Mechanics with Lorentz Covariance and $c \to \infty$ Limit,
Symmetry  {\bf 13} (2021) 22.
\bibitem{095}
O.C.W. Kong,
Quantum Origin of (Newtonian) Mass and and Galilean Relativity Symmetry,
Chin. J. Phys.  (2023) {\it to be published}.
\bibitem{081}
O.C.W. Kong, 
A Geometric Picture of Quantum Mechanics with Noncommutative Values for Observables, 
Results Phys.   {\bf 19} (2020) 103606.
\bibitem{Z} 
J.S. Zmuidzinas,  
Unitary Representations of the Lorentz Group on 4-Vector Manifolds,
 J. Math. Phys.  {\bf  7} (1966) 764-780.  
\bibitem{J}
J.E. Johnson, Position Operators and Proper Time in Relativistic Quantum Mechanics.
 Phys. Rev.   {\bf 181} (1969) 1755-1764.
\bibitem{T}
W.K. Tung,
{Group Theory in Physics}, World Scientific (1985).
\bibitem{TS}
M.A. Trump and W.C. Schieve, 
{Classical Relativistic Many-Body Dynamics},
Springer Science+Business Media (1999).
\bibitem{Ho}
L.P. Horwitz, 
{Relativistic Quantum Mechanics}
Springer Science+Business Media (2015).
\bibitem{F}
J.R. Fanchi, Parametrized Relativistic Quantum Theory, Kluwer Academic Publisher (1993).
\bibitem{ni} 
D.C. Currie, T.F. Jordan, and E.G.C. Sudarshan,  
Relativistic Invariance and Hamiltonian Theories of Interacting Particles,
Rev. Mod. Phys.  {\bf 35} (1963) 350-375 .
\bibitem{078}
O.C.W. Kong  and W.-Y. Liu,  
Noncommutative Coordinate Picture of the Quantum Phase Space,
Chin. J. Phys.  {\bf 77} (2022) 2881-2896.
\bibitem{qrf}
 F. Giacomini,  E. Castro-Ruiz, and \u{C}. Brukner, 
Quantum mechanics and the covariance of physical laws in quantum reference frames,
Nature Commun.  {\bf  10} (2019) 494. 
\bibitem{093}%
O.C.W. Kong, Quantum Frames of Reference  and the Noncommutative Values of Observables,
Results Phys.   {\bf 31} (2021) 105033.
\bibitem{PR} 
C. Piron and F. Reuse,
The Relativistic Two Body Problem,
Helv. Phys. Acta {\bf 48} (1975) 631-638. 
\bibitem{Jh}
O.D. Johns, Analytical Mechanics for Relativity and Quantum Mechanics, Oxford University Press (2005). 
\bibitem{M}
M. Maggiore, A Modern Introduction to Quantum Field Theory, Oxford University Press (2005).
\bibitem{086} %
O.C.W. Kong and  J. Payne,
Special Relativity and its Newtonian Limit from a Group Theoretical Perspective,
Symmetry {\bf 13} (2021) 1925.
\bibitem{R}
F. Rohrlich, Classical Charged Particles, World Scientific (2007).
\bibitem{FW}
J.A. Wheeler and R.F. Feynman,
Classical Electrodynamics in Terms of Direct Interparticle Action,
Rev. Mod. Phys. {\bf 21} (1949) 425-433.
 
\end{thebibliography}
\end{document}